\documentclass[preprint]{ptephy}

\preprintnumber{KYUSHU-HET-145, OIQP-14-13}

\usepackage{amssymb}
\usepackage{amsthm}
\usepackage{amsmath}
\usepackage{booktabs}

\numberwithin{equation}{section}

\usepackage{url}
\let\Gamma\varGamma

\begin{document}

\title{Large-$N$ limit of the gradient flow in the 2D $O(N)$ nonlinear sigma
model}

\author{%
\name{\fname{Hiroki} \surname{Makino}}{1},
\name{\fname{Fumihiko} \surname{Sugino}}{2}, and
\name{\fname{Hiroshi} \surname{Suzuki}}{1,\ast}
}

\address{%
\affil{1}{Department of Physics, Kyushu University, 6-10-1 Hakozaki,
Higashi-ku, Fukuoka, 812-8581, Japan}
\affil{2}{Okayama Institute for Quantum Physics, Kyoyama 1-9-1, Kita-ku,
Okayama 700-0015, Japan}
\email{hsuzuki@phys.kyushu-u.ac.jp}
}

\begin{abstract}
The gradient flow equation in the 2D $O(N)$ nonlinear sigma model with lattice
regularization is solved in the leading order of the $1/N$ expansion. By using
this solution, we analytically compute the thermal expectation value of a
lattice energy--momentum tensor defined through the gradient flow. The
expectation value reproduces thermodynamic quantities obtained by the standard
large-$N$ method. This analysis confirms that the above lattice
energy--momentum tensor restores the correct normalization automatically in the
continuum limit, in a system with a non-perturbative mass gap.
\end{abstract}
\subjectindex{B31, B32, B34, B38}
\maketitle

\section{Introduction}
\label{sec:1}
The Yang--Mills gradient flow or the Wilson flow~\cite{Luscher:2010iy} is a
powerful method to construct renormalized composite operators in gauge theory
(see~Ref.~\cite{Luscher:2013vga} for a recent review). This follows from the
fact that a local product of bare fields evolved by the gradient flow possesses
quite simple renormalization properties~\cite{Luscher:2011bx,Luscher:2013cpa}:
The multiplicative renormalization factor of the local product is determined
simply by the number of fermion (or generally matter) fields contained in the
local product; the flowed gauge field requires no multiplicative
renormalization. Furthermore, no infinite subtraction is needed. Since such a
renormalized operator is independent of regularization (after the parameter
renormalization), the gradient flow is expected to be quite useful in relating
physical quantities in continuum field theory and operators in lattice theory.

On the basis of this very general idea, a possible method to construct the
energy--momentum tensor on the lattice through the gradient flow was proposed
in~Ref.~\cite{Suzuki:2013gza}. This method was further investigated from a
somewhat different perspective in~Ref.~\cite{DelDebbio:2013zaa} and also
generalized in~Ref.~\cite{Makino:2014taa}. As well
recognized~\cite{Caracciolo:1988hc,Caracciolo:1989pt}, the construction of the
energy--momentum tensor on the lattice is quite involved because lattice
regularization breaks the translational invariance. The intention
of~Refs.~\cite{Suzuki:2013gza,Makino:2014taa} is that the constructed lattice
energy--momentum tensor restores the correct normalization and the conservation
law automatically in the continuum limit.

The construction in~Refs.~\cite{Suzuki:2013gza,Makino:2014taa} is based on very
natural assumptions, such as the existence of the energy--momentum tensor and
the renormalizability of the gradient flow in the non-perturbative level. Also,
the validity of the construction has been tested for thermodynamic quantities
in quenched QCD by using a Monte Carlo simulation~\cite{Asakawa:2013laa}. See
also Ref.~\cite{Kitazawa:2014uxa} for updated numerical results. However,
whether the conservation law is really restored in the non-perturbative level
is still to be carefully examined.

Under these situations, it must be instructive to consider a simpler system
that would allow a similar construction of the lattice energy--momentum tensor.
Mainly with this motivation, the gradient flow for the 2D $O(N)$ nonlinear
sigma model was investigated in~Ref.~\cite{Makino:2014sta}; an identical flow
equation has also been studied in~Ref.~\cite{Kikuchi:2014rla}.
In~Ref.~\cite{Makino:2014sta}, it was proven to all orders of perturbation
theory that the $N$-vector field evolved by the gradient flow requires no
multiplicative renormalization, a quite analogous property to the 4D gauge
field. Because of this renormalizability of the gradient flow and because of
the asymptotic freedom, one can imitate the construction of the lattice
energy--momentum tensor in~Refs.~\cite{Suzuki:2013gza,Makino:2014taa}. Then,
since the 2D $O(N)$ nonlinear sigma model is solvable in the $1/N$~expansion
(see, e.g., Ref.~\cite{Peskin:1995ev}), one naturally expects that the property
of the lattice energy--momentum tensor constructed through the gradient flow
can be investigated by utilizing this analytical method, without any systematic
errors associated with numerical study.

This is the main intention of the present paper: We test the construction of
the lattice energy--momentum tensor in~Ref.~\cite{Makino:2014sta} by using the
$1/N$~expansion. For this, we first recapitulate the well known large-$N$
solution of the 2D $O(N)$ nonlinear sigma model that exhibits a
non-perturbative mass gap (Sect.~\ref{sec:2}). Next, we solve the gradient flow
equation in the leading order of the $1/N$~expansion (Sect.~\ref{sec:3}). We
could not find a solution in the sub-leading order of the $1/N$~expansion. This
is unfortunate, because in the leading order of the $1/N$~expansion all
correlation functions factorize into one-point functions, while the test of the
conservation law of the energy--momentum tensor requires nontrivial multi-point
functions. Still, we can exactly compute one-point functions in the large-$N$
limit. For example, we can obtain a non-perturbative running coupling constant
by computing the vacuum expectation value of a composite operator analogous to
the ``energy density'' defined in~Ref.~\cite{Luscher:2010iy}
(Sect.~\ref{sec:4}). The one-point function of our energy--momentum tensor is
trivial in vacuum, but it becomes nontrivial if one considers the system at
finite temperature, as in~Ref.~\cite{Asakawa:2013laa}. In~Sect.~\ref{sec:5}, we
compute the expectation value of the energy--momentum tensor at finite
temperature in the large-$N$ limit. This expectation value is directly related
to thermodynamic quantities (the energy density and the pressure) of the
present system. We observe that the expectation value correctly reproduces
thermodynamic quantities directly computed by a standard statistical large-$N$
method given in~Appendix~\ref{sec:A}. In~Appendix~\ref{sec:B}, we illustrate
how a ``naive'' construction of the energy--momentum tensor on the lattice
fails to reproduce the correct answer. The present analytical test confirms
that the lattice energy--momentum tensor in~Ref.~\cite{Makino:2014sta} restores
the correct normalization in this system with a non-perturbative mass gap, at
least in the large-$N$ limit. The last section is devoted to the conclusion.

\section{Leading large-$N$ solution of  the 2D $O(N)$ nonlinear sigma model}
\label{sec:2}
The partition function of the 2D $O(N)$ nonlinear sigma model is given by
\begin{align}
   \mathcal{Z}
   &=\int
   \left[\prod_x\mathrm{d}\sigma(x)\,\right]\,
   \left[\prod_x\prod_{i=1}^N\mathrm{d}n^i(x)\right]\,
\notag\\
   &\qquad{}\times\exp\left(
   -\frac{1}{2\lambda_0}a^2\sum_x
   \left\{
   \partial_\mu n^i(x)\partial_\mu n^i(x)
   +\sigma(x)\left[n^i(x)n^i(x)-N\right]
   \right\}\right),
\label{eq:(2.1)}
\end{align}
where $\lambda_0$ is the bare 't~Hooft coupling constant, which is held fixed
in the large-$N$ limit. Throughout this paper, repeated Latin indices $i$, $j$,
\dots, are assumed to be summed over the integers from~$1$ to~$N$.
In~Eq.~\eqref{eq:(2.1)}, we assume lattice regularization with the lattice
spacing~$a$ and $\partial_\mu$ denotes the forward difference operator. To
apply the $1/N$ expansion (see, e.g., Ref.~\cite{Peskin:1995ev}), one first
integrates over the $N$-vector field~$n^i(x)$, to yield
\begin{align}
   \mathcal{Z}
   =\int
   \left[\prod_x\mathrm{d}\sigma(x)\,\right]\,
   \exp\left\{
   \frac{N}{2\lambda_0}a^2\sum_x\sigma(x)
   -\frac{N}{2}\ln\det\left[
   -\partial_\mu^*\partial_\mu+\sigma(x)
   \right]\right\},
\label{eq:(2.2)}
\end{align}
where $\partial_\mu^*$ denotes the backward difference operator. Then, since the
exponent is proportional to~$N$, for large $N$, the integral over the auxiliary
field~$\sigma(x)$ can be evaluated by the saddle point method. Assuming that
the saddle point is independent of~$x$, $\sigma(x)=\sigma$, it is given by the
gap equation,
\begin{equation}
   \frac{1}{\lambda_0}=\int_p\frac{1}{\Hat{p}^2+\sigma},\qquad
   \int_p\equiv\int_{-\pi/a}^{\pi/a}\frac{\mathrm{d}^2p}{(2\pi)^2},
\label{eq:(2.3)}
\end{equation}
where
\begin{equation}
   \Hat{p}^2\equiv\sum_\mu\Hat{p}_\mu\Hat{p}_\mu,\qquad
   \Hat{p}_\mu\equiv\frac{2}{a}\sin\left(\frac{1}{2}ap_\mu\right). 
\label{eq:(2.4)}
\end{equation}
An explicit momentum integration yields
\begin{equation}
   \frac{1}{\lambda_0}=\int_p\frac{1}{\Hat{p}^2+\sigma}
   \stackrel{a\to0}{\to}
   \frac{1}{4\pi}\left[-\ln(a^2\sigma)+5\ln2\right].
\label{eq:(2.5)}
\end{equation}

In the present problem, we may equally adopt dimensional regularization~(DR),
by setting the spacetime dimension~$D=2-\epsilon$. With this regularization,
the associated bare coupling constant~$\lambda_0^{\text{DR}}$ is renormalized as
\begin{equation}
   \lambda_0^{\text{DR}}=\mu^\epsilon\lambda Z,
\label{eq:(2.6)}
\end{equation}
with the renormalization scale~$\mu$. The gap equation is obtained
as~Eq.~\eqref{eq:(2.3)} and one has
\begin{equation}
   \frac{1}{\lambda_0^{\text{DR}}}
   =   \frac{1}{\mu^\epsilon\lambda Z}
   =\int\frac{\mathrm{d}^Dp}{(2\pi)^D}\,
   \frac{1}{p^2+\sigma}
   \stackrel{D\to2}{\to}
   \frac{1}{2\pi}\left[
   \frac{1}{\epsilon}
   -\frac{1}{2}\ln\left(\frac{e^\gamma\sigma}{4\pi}\right)\right],
\label{eq:(2.7)}
\end{equation}
where $\gamma$ is the Euler constant. From this expression, we can deduce the
exact renormalization constant in the minimal subtraction (MS) scheme,
\begin{equation}
   Z^{-1}=1+\frac{\lambda}{2\pi}\frac{1}{\epsilon},
\label{eq:(2.8)}
\end{equation}
and correspondingly the exact beta function,
\begin{equation}
   \beta\equiv\left.\mu\frac{\partial}{\partial\mu}\lambda
   \right|_{\text{$\lambda_0^{\text{DR}}$ fixed}}
   =-\epsilon\lambda-\frac{\lambda^2}{2\pi}.
\label{eq:(2.9)}
\end{equation}
Then, from Eq.~\eqref{eq:(2.7)}, we have
\begin{equation}
   \sigma=4\pi\mathrm{e}^{-\gamma}\mu^2\mathrm{e}^{-4\pi/\lambda}
   =4\pi\mathrm{e}^{-\gamma}\Lambda^2,\qquad
   \Lambda\equiv\mu\mathrm{e}^{-2\pi/\lambda},
\label{eq:(2.10)}
\end{equation}
in terms of the renormalized 't~Hooft coupling~$\lambda$ in the MS scheme.
Here, we have introduced the renormalization-group invariant scale
parameter~$\Lambda$ in the MS scheme. Going back to~Eq.~\eqref{eq:(2.1)}, the
saddle point value~$\sigma$ provides the mass gap for the $N$-vector field.
This mass gap is non-perturbative, as the dependence of~$\sigma$ on the
coupling constant~$\lambda$ shows.

\section{Leading large-$N$ solution of the gradient flow equation}
\label{sec:3}
Following Refs.~\cite{Makino:2014sta,Kikuchi:2014rla}, we consider the flow
equation in the $O(N)$ nonlinear sigma model defined by\footnote{Note that the
normalization of the $N$-vector field is different from that
of~Ref.~\cite{Makino:2014sta} by the factor~$1/\sqrt{N}$.}
\begin{equation}
   \partial_t n^i(t,x)
   =\partial_\mu^*\partial_\mu n^i(t,x)
   -\frac{1}{N}n^j(t,x)\partial_\mu^*\partial_\mu n^j(t,x)n^i(t,x),
\label{eq:(3.1)}
\end{equation}
where $t$ is the flow time and the initial value at~$t=0$ is given by the
$N$-vector field in the original $O(N)$ nonlinear sigma model,
\begin{equation}
   n^i(t=0,x)=n^i(x),
\label{eq:(3.2)}
\end{equation}
that is subject to the functional integral~\eqref{eq:(2.1)}. In this
expression, again, we are assuming lattice regularization in the
$x$~directions. To make the counting of the order of~$1/N$ easier, we render
the flow equation~\eqref{eq:(3.1)} linear in~$n^i(t,x)$ by introducing a new
variable~$\sigma(t,x)$ as,
\begin{align}
   \partial_t n^i(t,x)
   &=\partial_\mu^*\partial_\mu n^i(t,x)-\sigma(t,x)n^i(t,x),
\label{eq:(3.3)}
\\
   \sigma(t,x)&=\frac{1}{N}n^j(t,x)\partial_\mu^*\partial_\mu n^j(t,x).
\label{eq:(3.4)}
\end{align}
Note that the second relation does not contain the flow-time derivative. Then
Eq.~\eqref{eq:(3.3)} can be formally solved as
\begin{equation}
   n^i(t,x)=a^2\sum_y
   \left[
   K_t(x-y)n^i(y)
   -\int_0^t\mathrm{d}s\,K_{t-s}(x-y)\sigma(s,y)n^i(s,y)
   \right],
\label{eq:(3.5)}
\end{equation}
where
\begin{equation}
   K_t(x)\equiv\int_p\mathrm{e}^{ipx}\,\mathrm{e}^{-t\Hat{p}^2}
\label{eq:(3.6)}
\end{equation}
is the heat kernel with lattice regularization. The heat kernel satisfies
$\partial_t K_t(x)=\partial_\mu^*\partial_\mu K_t(x)$
and~$K_0(x)=\delta_{x,0}/a^2$. By iteratively solving Eq.~\eqref{eq:(3.5)}, we
can express the flowed field~$n^i(t,x)$ in terms of the initial value~$n^i(y)$
and~$\sigma(s,z)$ at intermediate flow times as
\begin{align}
   n^i(t,x)&=\sum_{m=0}^\infty(-1)^m
   a^2\sum_ya^2\sum_{z_1}a^2\sum_{z_2}\dotsm a^2\sum_{z_m}
\notag\\
   &\qquad{}
   \times
   \int_0^t\mathrm{d}s_1\,\int_0^{s_1}\mathrm{d}s_2\dotsm
   \int_0^{s_{m-1}}\mathrm{d}s_m\,
   \sigma(s_1,z_1)\sigma(s_2,z_2)\dotsm\sigma(s_m,z_m)
\notag\\
   &\qquad\qquad{}
   \times K_{t-s_1}(x-z_1)K_{s_1-s_2}(z_1-z_2)\dotsm K_{s_{m-1}-s_m}(z_{m-1}-z_m) 
\notag\\
   &\qquad\qquad\qquad\qquad{}
   \times K_{s_m}(z_m-y)n^i(y).
\label{eq:(3.7)}
\end{align}

\begin{figure}[ht]
\begin{center}
\includegraphics[scale=0.5,clip]{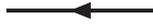}
\caption{Diagrammatic representation of the heat kernel~\eqref{eq:(3.6)}.}
\label{fig:1}
\end{center}
\end{figure}

\begin{figure}[ht]
\begin{center}
\includegraphics[scale=0.5,clip]{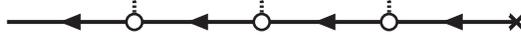}
\caption{The $m=3$~term in the solution~\eqref{eq:(3.7)}. The cross denotes the
$N$-vector field at zero flow time, $n^i(y)$.}
\label{fig:2}
\end{center}
\end{figure}

\begin{figure}[ht]
\begin{center}
\includegraphics[scale=0.5,clip]{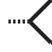}
\caption{Diagrammatic representation of the equality~\eqref{eq:(3.4)}, which
is~$O(1/N)$.}
\label{fig:3}
\end{center}
\end{figure}

Diagrammatic representation of the above elements and expressions is
useful.\footnote{The present convention for the ``flow Feynman diagram'' is
quite different from that in~Ref.~\cite{Makino:2014sta}.}
In~Eq.~\eqref{eq:(3.7)}, the heat kernel~$K_t(x)$~\eqref{eq:(3.6)} connecting
two spacetime points is represented by an arrowed solid line
as~Fig.~\ref{fig:1}. An open circle denotes the interaction between the flowed
$N$-vector field and the auxiliary field~$\sigma(t,x)$, which is represented by
a short dotted line. A typical term in the solution~\eqref{eq:(3.7)} is thus
represented as~Fig.~\ref{fig:2}, where the $N$-vector field at the zero flow
time, $n^i(y)$, is represented by the cross. The equality~\eqref{eq:(3.4)} is,
on the other hand, represented as~Fig.~\ref{fig:3}, where two short solid lines
represent two $N$-vector fields in the right-hand side of~Eq.~\eqref{eq:(3.4)}.
Note that Eq.~\eqref{eq:(3.4)} and thus the symbol in~Fig.~\ref{fig:3} carry
the factor~$1/N$. 

\begin{figure}[ht]
\begin{center}
\includegraphics[scale=0.5,clip]{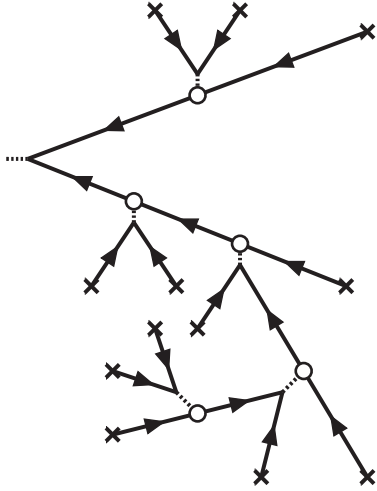}
\caption{$\sigma(t,x)$ in terms of the zero flow-time field~$n^i(x)$.}
\label{fig:4}
\end{center}
\end{figure}

We may now substitute the solution~\eqref{eq:(3.7)} in the
equality~\eqref{eq:(3.4)} to express the auxiliary field~$\sigma(t,x)$ in terms
of the initial value~$n^i(y)$. This process can be diagrammatically represented
as~Fig.~\ref{fig:4}.

\begin{figure}[ht]
\begin{center}
\includegraphics[scale=0.5,clip]{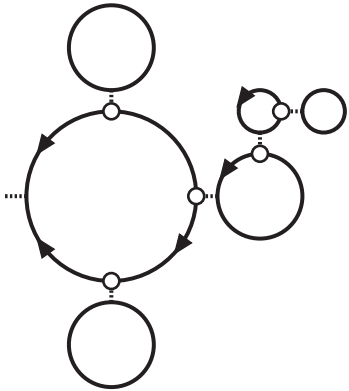}
\caption{$\langle\sigma(t,x)\rangle$ in the leading order of the $1/N$
expansion; this diagram is obtained by taking the contraction of the $n^i(y)$
in~Fig.~\ref{fig:4}.}
\label{fig:5}
\end{center}
\end{figure}

So far, everything concerns the solution to the deterministic differential
equation~\eqref{eq:(3.1)}. Let us now take into account the quantum effect,
i.e., the fact that the initial value~$n^i(y)$ is subject to the quantum
average~\eqref{eq:(2.1)}. In the leading order of the $1/N$~expansion, the
integration over the auxiliary field~$\sigma(x)$ in~Eq.~\eqref{eq:(2.1)} is
approximated by the value at the saddle point, $\sigma(x)=\sigma$. Then, since
the action is quadratic in~$n^i(x)$, the quantum average produces contractions
of~$n^i(x)$~fields by the free massive propagator with the mass~$\sigma$. In
terms of the diagrammatic representation above, this amounts to taking the
contraction of all crosses in all possible ways. Let us consider these
contractions for~$\sigma(t,x)$ in~Fig.~\ref{fig:4}. In this diagram, recalling
that the vertex in~Fig.~\ref{fig:3} carries the factor~$1/N$ and noting that
each closed loop of the $N$-vector field gains the factor~$N$, it is obvious
that the leading large-$N$ contribution to the quantum average
of~$\sigma(t,x)$, denoted by~$\langle\sigma(t,x)\rangle$, is given by a diagram
such as~Fig.~\ref{fig:5} in which each closed loop contains only one vertex
in~Fig.~\ref{fig:3}; overall, this is a quantity of~$O(N^0)$.\footnote{In the
diagrammatic representation, we adopt a rule~\cite{Luscher:2011bx} that arrows
are removed when end points of arrowed lines are contracted.} The topology of
diagrams in the leading order in the $1/N$~expansion is thus identical to that
of the leading order diagrams in the conventional $1/N$~expansion of the
$N$-vector model (the so-called ``cactus'' diagrams). To calculate sub-leading
orders of~$1/N$, we have to find not only the one-point function but also the
(connected) higher-point functions of~$\sigma(t,x)$, whose systematic treatment
is left as a future subject.

\begin{figure}[ht]
\begin{center}
\includegraphics[scale=0.5,clip]{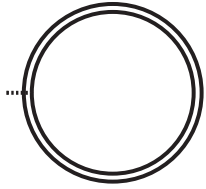}
\caption{Figure~\ref{fig:5} in terms of the dressed
propagator~\eqref{eq:(3.11)} (the doubled line).}
\label{fig:6}
\end{center}
\end{figure}

In a similar manner, it is easy to see that, in the leading order in the
$1/N$~expansion, a correlation function of generic operators containing
$\sigma(t,x)$ and~$n^i(t,x)$ fields factorizes into the product of the
expectation value~$\langle\sigma(t,x)\rangle$ and correlation functions of the
$n^i(t,x)$; this is nothing but the large-$N$ factorization. Then, since
$\langle\sigma(t,x)\rangle$ is independent of the spacetime position~$x$ (the
external momentum in~Fig.~\ref{fig:5} is zero), we can set $\sigma(s,z)$
in~Eq.~\eqref{eq:(3.7)} constant in spacetime,
$\sigma(s,z)\to\langle\sigma(s)\rangle$.\footnote{Note that, since there is no
translational invariance in the flow-time direction (the zero flow time is a
very special point), we cannot assume that $\langle\sigma(s)\rangle$ is
independent of~$s$. In fact, we will shortly see that $\langle\sigma(s)\rangle$
possesses nontrivial $s$~dependence.} Then, noting the relation
\begin{equation}
   a^2\sum_zK_{t-u}(x-z)K_{u-s}(z-y)=K_{t-s}(x-y),
\label{eq:(3.8)}
\end{equation}
we have a compact expression for~Eq.~\eqref{eq:(3.7)},
\begin{equation}
   n^i(t,x)=\mathrm{e}^{-\int_0^t\mathrm{d}s\,\sigma(s)}a^2\sum_yK_t(x-y)n^i(y),
\label{eq:(3.9)}
\end{equation}
where we have written $\sigma(s)\equiv\left\langle\sigma(s)\right\rangle$ for
notational simplicity. The propagator between the flowed $N$-vector fields is
then obtained by contracting $n^i(y)$ in~Eq.~\eqref{eq:(3.9)} by the propagator
in the large-$N$ limit:
\begin{equation}
   \left\langle n^i(x)n^j(y)\right\rangle
   =\delta^{ij}\lambda_0
   \int_p\mathrm{e}^{ip(x-y)}\frac{1}{\Hat{p}^2+\sigma}.
\label{eq:(3.10)}
\end{equation}
This yields
\begin{equation}
   \left\langle n^i(t,x)n^j(s,y)\right\rangle
   =\delta^{ij}\mathrm{e}^{-\int_0^t\mathrm{d}u\,\sigma(u)}
   \mathrm{e}^{-\int_0^s\mathrm{d}v\,\sigma(v)}
   \lambda_0
   \int_p\mathrm{e}^{ip(x-y)}
   \frac{\mathrm{e}^{-(t+s)\Hat{p}^2}}{\Hat{p}^2+\sigma}.
\label{eq:(3.11)}
\end{equation}
In terms of this ``dressed propagator'', the expectation
value~$\langle\sigma(t)\rangle$ is given from~Eq.~\eqref{eq:(3.4)} by
\begin{align}
   \sigma(t)
   &=\left\langle\frac{1}{N}
   n^i(t,x)\partial_\mu^*\partial_\mu n^i(t,x)\right\rangle
\notag\\
   &=\mathrm{e}^{-2\int_0^t\mathrm{d}s\,\sigma(s)}\lambda_0
   \int_p\frac{-\Hat{p}^2}{\Hat{p}^2+\sigma}\mathrm{e}^{-2t\Hat{p}^2}.
\label{eq:(3.12)}
\end{align}
This self-consistency condition is schematically represented
as~Fig.~\ref{fig:6}.

Now we solve the self-consistency condition for~$\sigma(t)$,
Eq.~\eqref{eq:(3.12)}. For this, we introduce
\begin{equation}
   \Sigma(t)=\int_0^t\mathrm{d}s\,\sigma(s),
\label{eq:(3.13)}
\end{equation}
and write Eq.~\eqref{eq:(3.12)} as
\begin{equation}
   \mathrm{e}^{2\Sigma(t)}\frac{\mathrm{d}\Sigma(t)}{\mathrm{d}t}
   =\lambda_0\int_p\frac{-\Hat{p}^2}{\Hat{p}^2+\sigma}\mathrm{e}^{-2t\Hat{p}^2}.
\label{eq:(3.14)}
\end{equation}
As far as lattice regularization is understood, the momentum integration in the
right-hand side is regular even at~$t=0$ and we may integrate both sides of the
above relation over~$t$ from $t=0$ to some prescribed value. In this way, we
have
\begin{equation}
   \Sigma(t)=\frac{1}{2}\ln
   \left(\lambda_0\int_p\frac{\mathrm{e}^{-2t\Hat{p}^2}}{\Hat{p}^2+\sigma}
   \right),
\label{eq:(3.15)}
\end{equation}
where we have used the saddle point condition~\eqref{eq:(2.3)}. Substituting
this back into~Eq.~\eqref{eq:(3.12)} leads to
\begin{equation}
   \sigma(t)=\sigma
   -\frac{\displaystyle\int_p\mathrm{e}^{-2t\Hat{p}^2}}
   {\displaystyle\int_p\frac{\mathrm{e}^{-2t\Hat{p}^2}}{\Hat{p}^2+\sigma}}.
\label{eq:(3.16)}
\end{equation}
As far as~$t>0$, the integrals are well convergent and we may send $a\to0$ to
have a definite continuum limit. Thus, for~$t>0$, we obtain
\begin{equation}
   \sigma(t)
   \stackrel{a\to0}{\to}
   \sigma-
   \frac{1}{2t\mathrm{e}^{2\sigma t}\Gamma(0,2\sigma t)}
   \stackrel{t\to0}{\to}
   \sigma+\frac{1}{2t\ln(2\mathrm{e}^\gamma\sigma t)}
   \left[1-2\sigma t+O(t/\ln t)\right],
\label{eq:(3.17)}
\end{equation}
where $\Gamma(z,p)$ is the incomplete gamma function. Here, the order of the
two limits is very important. Our construction of the energy--momentum tensor
on the basis of the gradient flow relies on a universality, which is ensured if
the flow time is fixed and ultraviolet regularization is removed. Thus, we
should first take the continuum limit while keeping the flow time finite; we
then consider the small flow-time limit. Also, using Eqs.~\eqref{eq:(3.12)}
and~\eqref{eq:(3.16)}, for~$t>0$ we have
\begin{equation}
   \mathrm{e}^{-2\int_0^t\mathrm{d}s\,\sigma(s)}\lambda_0
   =\frac{\displaystyle1}
   {\displaystyle\int_p\frac{\mathrm{e}^{-2t\Hat{p}^2}}{\Hat{p}^2+\sigma}}
   \stackrel{a\to0}{\to}
   \frac{4\pi}{\mathrm{e}^{2\sigma t}\Gamma(0,2\sigma t)}
   \stackrel{t\to0}{\to}
   -\frac{4\pi}
   {\ln(2\mathrm{e}^\gamma\sigma t)}
   \left[1-2\sigma t+O(t/\ln t)\right].
\label{eq:(3.18)}
\end{equation}
The dressed propagator~\eqref{eq:(3.11)} with this prefactor provides the
solution of the gradient flowed system at the leading order in the large-$N$
limit.

\section{Non-perturbative running coupling in the large-$N$ limit}
\label{sec:4}
Since the expectation value,
\begin{equation}
   \lambda_R(1/\sqrt{8t})
   \equiv 16\pi t\left\langle E(t,x)\right\rangle,
\label{eq:(4.1)}
\end{equation}
where
\begin{equation}
   E(t,x)\equiv\frac{1}{2}\frac{1}{N}\partial_\mu n^i(t,x)\partial_\mu n^i(t,x),
\label{eq:(4.2)}
\end{equation}
is a renormalized quantity~\cite{Makino:2014sta} that possesses the
perturbative expansion, $16\pi t\langle E(t,x)\rangle=\lambda_0+\dotsb$, it can
be used as a non-perturbative definition of the running coupling constant at
the renormalization scale~$1/\sqrt{8t}$~\cite{Makino:2014sta}. This is
analogous to the non-perturbative running gauge coupling defined through the
``energy density operator''~\cite{Luscher:2010iy}.

From our large-$N$ solution in the previous section, we have
\begin{equation}
   \lambda_R(1/\sqrt{8t})
   =-8\pi t\sigma(t)
   \stackrel{a\to0}{\to}
   -8\pi\sigma t
   +\frac{4\pi}{\mathrm{e}^{2\sigma t}\Gamma(0,2\sigma t)},\qquad t>0.
\label{eq:(4.3)}
\end{equation}
This is a monotonically increasing function of~$t$ being consistent with the
fact that the exact beta function~\eqref{eq:(2.9)} is negative definite.

\section{Thermal expectation value of the lattice energy--momentum tensor}
\label{sec:5}
Following the general idea in~Refs.~\cite{Suzuki:2013gza,Makino:2014taa}, a
possible method using the gradient flow to construct a lattice energy--momentum
tensor for the $O(N)$ nonlinear sigma model has been
proposed~\cite{Makino:2014sta}. The intention in~Ref.~\cite{Makino:2014sta} is
to construct a lattice operator that restores the correct normalization and the
conservation law automatically in the continuum limit. It is thus quite
interesting to examine if the idea works (or not) by using the above exact
large-$N$ solution of the gradient flow. Unfortunately, at the leading order of
the $1/N$ expansion, any correlation function factorizes into one-point
functions of $O(N)$ invariant quantities. Thus, in the present paper, we can
consider only the one-point function of the energy--momentum tensor. Since we
define the energy--momentum tensor by subtracting the vacuum expectation value,
\begin{equation}
   \left\{T_{\mu\nu}\right\}_R(x)
   \equiv T_{\mu\nu}(x)-\left\langle T_{\mu\nu}(x)\right\rangle,
\label{eq:(5.1)}
\end{equation}
the one-point function is trivial in the vacuum. The one-point function of the
energy--momentum tensor is quite interesting, however, if we consider the
system at finite temperature, as in~Ref.~\cite{Asakawa:2013laa}. Thus, let us
consider the expectation value of the energy--momentum tensor at finite
temperature. The construction in~Ref.~\cite{Makino:2014sta} adopted in the
present large-$N$ limit reads
\begin{align}
   &\left\{T_{\mu\nu}\right\}_R(x)
\notag\\
   &=\lim_{t\to0}\lim_{a\to0}\biggl\{
   c_1(t)\left[
   \partial_\mu n^i(t,x)\partial_\nu n^i(t,x)
   -\frac{1}{2}\delta_{\mu\nu}\partial_\rho n^i(t,x)\partial_\rho n^i(t,x)
   \right]
\notag\\
   &\qquad\qquad\qquad{}
   +c_2(t)\left[
   \frac{1}{2}
   \delta_{\mu\nu}\partial_\rho n^i(t,x)\partial_\rho n^i(t,x)
   -\left\langle
   \frac{1}{2}
   \delta_{\mu\nu}\partial_\rho n^i(t,x)\partial_\rho n^i(t,x)
   \right\rangle
   \right]\biggr\},
\label{eq:(5.2)}
\end{align}
where the coefficients are given by
\begin{equation}
   c_1(t)=\frac{1}{\Bar{\lambda}(1/\sqrt{8t})}
   -\frac{1}{4\pi}\ln\pi+O(\Bar{\lambda}),\qquad
   c_2(t)=\frac{1}{4\pi}
   -\frac{1}{(4\pi)^2}\Bar{\lambda}(1/\sqrt{8t})+O(\Bar{\lambda}^2),
\label{eq:(5.3)}
\end{equation}
and
\begin{align}
   \Bar{\lambda}(q)
   =-\frac{4\pi}{\ln(\Lambda^2/q^2)}
\label{eq:(5.4)}
\end{align}
is the running coupling constant at the renormalization scale~$q$. From the
expressions in~Ref.~\cite{Makino:2014sta} (with the normalization change
$n^i(t,x)\to n^i(t,x)/\sqrt{N}$), these expressions are obtained by
setting~$g^2=\lambda/N$ and taking $N\to\infty$.

The expectation value of the energy--momentum tensor at finite temperature,
\begin{equation}
   \left\langle\left\{T_{\mu\nu}\right\}_R(x)\right\rangle_\beta,
\label{eq:(5.5)}
\end{equation}
where $\beta$ is the inverse temperature, is then obtained by contracting
$n^i(t,x)$ by the dressed propagator~\eqref{eq:(3.11)} with the periodic
boundary condition in the Euclidean time direction~$x_0$; the time component of
the momentum in~Eq.~\eqref{eq:(3.11)} is thus quantized to the Matsubara
frequency:
\begin{equation}
   p_0=\omega_n\equiv\frac{2\pi n}{\beta},\qquad n\in\mathbb{Z}.
\label{eq:(5.6)}
\end{equation}
Thus, for instance, we have
\begin{align}
   &\left\langle\partial_0 n^i(t,x)\partial_0 n^i(t,x)\right\rangle_\beta
\notag\\
   &=N\mathrm{e}^{-2\int_0^t\mathrm{d}s\,\sigma_\beta(s)}\lambda_0
   \frac{1}{\beta}\sum_{-\pi/a<\omega_n<\pi/a}
   \int_{-\pi/a}^{\pi/a}\frac{\mathrm{d}p_1}{2\pi}\,
   \frac{\Hat{\omega_n}^2}{\Hat{\omega_n}^2+\Hat{p_1}^2+\sigma_\beta}
   \mathrm{e}^{-2t(\Hat{\omega_n}^2+\Hat{p_1}^2)},
\label{eq:(5.7)}
\end{align}
where $\sigma_\beta(s)$ is the flow-time-dependent auxiliary field at finite
temperature that fulfills a finite temperature counterpart
of~Eq.~\eqref{eq:(3.18)}:
\begin{equation}
   \mathrm{e}^{-2\int_0^t\mathrm{d}s\,\sigma_\beta(s)}\lambda_0
   =\frac{\displaystyle1}
   {\displaystyle
   \frac{1}{\beta}\sum_{-\pi/a<\omega_n<\pi/a}
   \int_{-\pi/a}^{\pi/a}\frac{\mathrm{d}p_1}{2\pi}
   \frac{\mathrm{e}^{-2t(\Hat{\omega_n}^2+\Hat{p_1}^2)}}
   {\Hat{\omega_n}^2+\Hat{p_1}^2+\sigma_\beta}}.
\label{eq:(5.8)}
\end{equation}
On the other hand, $\sigma_\beta$ is the saddle point value of the auxiliary
field at finite temperature which is given by
\begin{equation}
   \frac{1}{\lambda_0}
   =\frac{1}{\beta}\sum_{-\pi/a<\omega_n<\pi/a}
   \int_{-\pi/a}^{\pi/a}\frac{\mathrm{d}p_1}{2\pi}
   \frac{1}{\Hat{\omega_n}^2+\Hat{p_1}^2+\sigma_\beta}.
\label{eq:(5.9)}
\end{equation}

Now, in expressions such as Eqs.~\eqref{eq:(5.7)} and~\eqref{eq:(5.8)}, the sum
and the integral are well convergent for~$t>0$ because of the Gaussian damping
factor. Thus we may simply remove lattice regularization in those expressions
to yield regularization-independent expressions such as
\begin{align}
   \left\langle\partial_0 n^i(t,x)\partial_0 n^i(t,x)\right\rangle_\beta
   &=N\mathrm{e}^{-2\int_0^t\mathrm{d}s\,\sigma_\beta(s)}\lambda_0
   \frac{1}{\beta}\sum_{n=-\infty}^\infty
   \int\frac{\mathrm{d}p_1}{2\pi}\,
   \frac{\omega_n^2}{\omega_n^2+p_1^2+\sigma_\beta}
   \mathrm{e}^{-2t(\omega_n^2+p_1^2)},
\label{eq:(5.10)}
\\
   \left\langle\partial_1 n^i(t,x)\partial_1 n^i(t,x)\right\rangle_\beta
   &=N\mathrm{e}^{-2\int_0^t\mathrm{d}s\,\sigma_\beta(s)}\lambda_0
   \frac{1}{\beta}\sum_{n=-\infty}^\infty
   \int\frac{\mathrm{d}p_1}{2\pi}\,
   \frac{p_1^2}{\omega_n^2+p_1^2+\sigma_\beta}
   \mathrm{e}^{-2t(\omega_n^2+p_1^2)}
\label{eq:(5.11)}
\end{align}
and
\begin{equation}
   \mathrm{e}^{-2\int_0^t\mathrm{d}s\,\sigma_\beta(s)}\lambda_0
   =\frac{\displaystyle1}
   {\displaystyle
   \frac{1}{\beta}\sum_{n=-\infty}^\infty
   \int\frac{\mathrm{d}p_1}{2\pi}
   \frac{1}{\omega_n^2+p_1^2+\sigma_\beta}\mathrm{e}^{-2t(\omega_n^2+p_1^2)}}.
\label{eq:(5.12)}
\end{equation}
These clearly illustrate the ``UV finiteness'' of the gradient flow: Any
correlation function of the flowed $N$-vector field in terms of the
renormalized coupling is UV finite without the wave function
renormalization~\cite{Makino:2014sta}.\footnote{From Eqs.~\eqref{eq:(2.5)}
and~\eqref{eq:(5.9)}, one sees that the ratio between~$\sigma_\beta$
and~$\sigma$ is a UV convergent quantity that is independent of the
regularization; the explicit relation is given by~\eqref{eq:(5.18)}. Thus, as
long as we renormalize the bare coupling constants $\lambda_0$
and~$\lambda_0^{\text{DR}}$ so that $\sigma$ in~Eqs.~\eqref{eq:(2.5)}
and~\eqref{eq:(2.7)} are identical, $\sigma_\beta$ defined
in~Eq.~\eqref{eq:(5.9)} through lattice regularization and $\sigma_\beta$
defined in~Eq.~\eqref{eq:(A2)} through dimensional regularization are
identical; $\sigma_\beta$ is of course finite after the renormalization.} It is
the basic idea for the construction of the lattice energy--momentum tensor
in~Refs.~\cite{Suzuki:2013gza,Makino:2014taa,Makino:2014sta} that the continuum
limit~$a\to0$ of a lattice composite operator of the flowed field reduces to a
regularization-independent expression. Thus, we have observed that the
continuum limit~$a\to0$ in~Eq.~\eqref{eq:(5.2)} can be almost trivially taken.
Next, to consider the small flow-time limit~$t\to0$ in~Eq.~\eqref{eq:(5.2)}, we
estimate the sum and the integral appearing in the above expressions
for~$t\to0$. This can be accomplished by noting the Poisson resummation
formula,
\begin{equation}
   \sum_{n=-\infty}^\infty\mathrm{e}^{-\alpha n^2}
   =\sqrt{\frac{\pi}{\alpha}}\sum_{n=-\infty}^\infty\mathrm{e}^{-\pi^2n^2/\alpha},
\label{eq:(5.13)}
\end{equation}
and, after some calculation, we have the following asymptotic expansions
for~$t\to0$:
\begin{align}
   &\frac{1}{\beta}\sum_{n=-\infty}^\infty\int\frac{\mathrm{d}p_1}{2\pi}\,
   \mathrm{e}^{-2t(\omega_n^2+p_1^2)}
   \sim\frac{1}{4\pi}\frac{1}{2t},
\label{eq:(5.14)}
\\
   &\frac{1}{\beta}\sum_{n=-\infty}^\infty\int\frac{\mathrm{d}p_1}{2\pi}\,
   \frac{1}{\omega_n^2+p_1^2+\sigma_\beta}\,
   \mathrm{e}^{-2t(\omega_n^2+p_1^2)}
\notag
\\
   &\sim-\frac{1}{4\pi}
   \ln(2\mathrm{e}^\gamma\sigma_\beta t)
   +\frac{1}{\pi}
   \sum_{n=1}^\infty K_0(\beta\sqrt{\sigma_\beta}n)
\notag\\
   &\qquad{}
   -\frac{1}{2\pi}\sigma_\beta t\left[\ln(2\mathrm{e}^\gamma\sigma_\beta t)-1
   \right]
   +\frac{2}{\pi}\sigma_\beta t
   \sum_{n=1}^\infty K_0(\beta\sqrt{\sigma_\beta}n)
   +O(t^2\ln t),
\label{eq:(5.15)}
\\
   &\frac{1}{\beta}\sum_{n=-\infty}^\infty\int\frac{\mathrm{d}p_1}{2\pi}\,
   \frac{\omega_n^2}{\omega_n^2+p_1^2+\sigma_\beta}\,
   \mathrm{e}^{-2t(\omega_n^2+p_1^2)}
\notag\\
   &\sim\frac{1}{8\pi}
   \left[\frac{1}{2t}
   +\sigma_\beta\ln(2\mathrm{e}^\gamma\sigma_\beta t)\right]
   +\frac{1}{\pi}\sigma_\beta
   \sum_{n=1}^\infty
   \left[
   \frac{1}{\beta\sqrt{\sigma_\beta}n}K_1(\beta\sqrt{\sigma_\beta}n)
   -K_2(\beta\sqrt{\sigma_\beta}n)
   \right]
   +O(t\ln t),
\label{eq:(5.16)}
\\
   &\frac{1}{\beta}\sum_{n=-\infty}^\infty\int\frac{\mathrm{d}p_1}{2\pi}\,
   \frac{p_1^2}{\omega_n^2+p_1^2+\sigma_\beta}\,
   \mathrm{e}^{-2t(\omega_n^2+p_1^2)}
\notag\\
   &\sim\frac{1}{8\pi}
   \left[\frac{1}{2t}+\sigma_\beta\ln(2\mathrm{e}^\gamma\sigma_\beta t)
   \right]
   +\frac{1}{\pi}\sigma_\beta
   \sum_{n=1}^\infty
   \frac{1}{\beta\sqrt{\sigma_\beta}n}K_1(\beta\sqrt{\sigma_\beta}n)
   +O(t\ln t),
\label{eq:(5.17)}
\end{align}
where $K_n(z)$ denotes the modified Bessel function of the $n$th kind. At this
stage, we note the following relation:
\begin{equation}
   -\frac{1}{4\pi}
   \ln(2\mathrm{e}^\gamma\sigma_\beta t)
   +\frac{1}{\pi}
   \sum_{n=1}^\infty K_0(\beta\sqrt{\sigma_\beta}n)
   =-\frac{1}{4\pi}
   \ln(2\mathrm{e}^\gamma\sigma t),
\label{eq:(5.18)}
\end{equation}
which can be obtained by comparing two gap equations, Eqs.~\eqref{eq:(2.7)}
and~\eqref{eq:(A7)}. By using this in~Eq.~\eqref{eq:(5.15)} and then
in~Eq.~\eqref{eq:(5.12)}, we find the asymptotic behavior of the prefactor
for~$t\to0$:
\begin{equation}
   \mathrm{e}^{-2\int_0^t\mathrm{d}s\,\sigma_\beta(s)}\lambda_0
   \sim-\frac{4\pi}{\ln(2\mathrm{e}^\gamma\sigma t)}
   \left[1-2\sigma_\beta t+O(t/\ln t)\right].
\label{eq:(5.19)}
\end{equation}
Also, from Eqs.~\eqref{eq:(5.3)}, \eqref{eq:(5.4)}, and~\eqref{eq:(2.10)},
for~$t\to0$, 
\begin{equation}
   c_1(t)\sim-\frac{1}{4\pi}\ln(2\mathrm{e}^\gamma\sigma t)+O(1/\ln t),\qquad
   c_2(t)\sim\frac{1}{4\pi}
   \left[1+\frac{1}{\ln(2\mathrm{e}^\gamma\sigma t/\pi)}\right]+O(1/\ln^2t).
\label{eq:(5.20)}
\end{equation}

It is now straightforward to obtain the $t\to0$ limit in~Eq.~\eqref{eq:(5.2)}.
Noting that $\beta\to\infty$ and~$\sigma_\beta\to\sigma$ on the vacuum, we have
\begin{align}
   \left\langle\left\{T_{00}\right\}_R(x)\right\rangle_\beta
   &=-\frac{N}{8\pi}(\sigma_\beta-\sigma)
   -\frac{N}{2\pi}\sigma_\beta
   \sum_{n=1}^\infty K_2(\beta\sqrt{\sigma_\beta}n),
\label{eq:(5.21)}
\\
   \left\langle\left\{T_{11}\right\}_R(x)\right\rangle_\beta
   &=-\frac{N}{8\pi}(\sigma_\beta-\sigma)
   +\frac{N}{2\pi}\sigma_\beta
   \sum_{n=1}^\infty K_2(\beta\sqrt{\sigma_\beta}n),
\label{eq:(5.22)}
\\
   \left\langle\left\{T_{01}\right\}_R(x)\right\rangle_\beta
   &=0.
\label{eq:(5.23)}
\end{align}
In this calculation, one finds that $1/t$~singularities are canceled between
the expectation value at finite temperature and the vacuum expectation value,
and a finite small flow-time limit results.

The thermodynamic quantities, the energy density~$\varepsilon$ and the
pressure~$P$, are related to these expectation values of the energy--momentum
tensor as
\begin{equation}
   \varepsilon=-\left\langle\left\{T_{00}\right\}_R(x)\right\rangle_\beta\qquad
   \text{and}
   \qquad
   P=\left\langle\left\{T_{11}\right\}_R(x)\right\rangle_\beta.
\label{eq:(5.24)}
\end{equation}
In Appendix~\ref{sec:A}, we compute these thermodynamic quantities by the
standard large-$N$ method. We find that Eq.~\eqref{eq:(5.24)} with
Eqs.~\eqref{eq:(5.21)} and~\eqref{eq:(5.22)} correctly reproduces those
large-$N$ results.


\section{Conclusion}
\label{sec:6}
In the present paper, we solved the gradient flow equation for the 2D $O(N)$
nonlinear sigma model in the leading order of the large-$N$ expansion. By using
this solution, one can non-perturbatively compute one-point functions of $O(N)$
invariant composite operators made from the flowed $N$-vector field in the
large-$N$ limit. We computed a non-perturbative running coupling from the
expectation value of the ``energy density operator'' in which the flow time
gives the renormalization scale. We also computed the thermal expectation value
of the lattice energy--momentum tensor, which is defined by a small flow time
limit of composite operators of the flowed field~\cite{Makino:2014sta}. We
found that the small flow time limit can be taken as expected and the lattice
energy--momentum tensor correctly reproduces the thermodynamic quantities
obtained by the standard large-$N$ approximation. This result for the present
system with a non-perturbatively generated mass gap strongly supports the
correctness of the reasoning for the lattice energy--momentum tensor
in~Refs.~\cite{Suzuki:2013gza,Makino:2014taa,Makino:2014sta}.

Quite unfortunately, in the present work, we could not find the solution for
the gradient flow equation in the next-to-leading order of the large-$N$
expansion. If this solution is obtained, it will make possible the examination
of the conservation law of the lattice energy--momentum tensor. We hope to come
back to this problem in the near future.

We would like to thank Kengo Kikuchi for the discussion.
F.S. would like to thank the members of KIAS, especially Hyeonjoon Shin, for
their warm hospitality during his visit.
The work of F.S. and H.S. is supported in part by Grants-in-Aid for
Scientific Research 25400289 and~23540330, respectively.

\section*{Note added}
In a recent paper~\cite{Aoki:2014dxa}, some of the results presented in this
paper have been obtained independently.

\appendix
\section{Thermodynamics at large~$N$}
\label{sec:A}
In the large-$N$ limit, the free energy density of the 2D $O(N)$ nonlinear
sigma model at finite temperature is given by, as a natural generalization of
the zero-temperature expression~\eqref{eq:(2.2)},
\begin{equation}
   f(\beta)
   =-\frac{N}{2\lambda_0}\beta\sigma_\beta
   +\frac{N}{2}\sum_{n=-\infty}^\infty
   \int\frac{\mathrm{d}p_1}{2\pi}\,\ln(\omega_n^2+p_1^2+\sigma_\beta),\qquad
   \omega_n\equiv\frac{2\pi}{\beta}n,
\label{eq:(A1)}
\end{equation}
where $\sigma_\beta$ denotes the saddle point value of the auxiliary
field~$\sigma(x)$ at finite temperature which is given by the solution of the
finite temperature gap equation:
\begin{equation}
   \frac{1}{\lambda_0}
   =\frac{1}{\beta}\sum_{n=-\infty}^\infty\int\frac{\mathrm{d}p_1}{2\pi}
   \frac{1}{\omega_n^2+p_1^2+\sigma_\beta}.
\label{eq:(A2)}
\end{equation}

We regularize the formal expressions~\eqref{eq:(A1)} and~\eqref{eq:(A2)} by
using dimensional regularization. For this, we note the identity
\begin{equation}
   \frac{1}{\beta}\sum_{n=-\infty}^\infty F(\omega_n)
   =\sum_{n=-\infty}^\infty\int\frac{\mathrm{d}p_0}{2\pi}\,\mathrm{e}^{ip_0\beta n}
   F(p_0),
\label{eq:(A3)}
\end{equation}
and regularize Eq.~\eqref{eq:(A1)} as
\begin{equation}
   f(\beta)
   \equiv-\frac{N}{2\lambda_0^{\text{DR}}}\beta\sigma_\beta
   +\frac{N}{2}\beta\sum_{n=-\infty}^\infty
   \int\frac{\mathrm{d}^Dp}{(2\pi)^D}\,\mathrm{e}^{ip_0\beta n}
   \ln(p^2+\sigma_\beta),
\label{eq:(A4)}
\end{equation}
where $\lambda_0^{\text{DR}}$ is the bare coupling constant in dimensional
regularization appearing in~Eq.~\eqref{eq:(2.7)}, and Eq.~\eqref{eq:(A2)} as
\begin{equation}
   \frac{1}{\lambda_0^{\text{DR}}}
   =\sum_{n=-\infty}^\infty
   \int\frac{\mathrm{d}^Dp}{(2\pi)^D}\,\mathrm{e}^{ip_0\beta n}
   \frac{1}{p^2+\sigma_\beta}.
\label{eq:(A5)}
\end{equation}
In the second term of the right-hand side of~Eq.~\eqref{eq:(A4)}, only the
$n=0$ term requires regularization because the $n\neq0$ terms are Fourier
transformations and UV convergent. After the momentum integration, we have
\begin{equation}
   f(\beta)
   =-\frac{N}{2\lambda_0^{\text{DR}}}\beta\sigma_\beta
   +\frac{N}{4\pi}\beta\sigma_\beta
   \left\{\frac{1}{\epsilon}
   -\frac{1}{2}\left[\ln\left(\frac{\mathrm{e}^\gamma\sigma_\beta}{4\pi}\right)
   -1\right]\right\}
   -\frac{N}{\pi}\beta\sigma_\beta
   \sum_{n=1}^\infty\frac{K_1(\beta\sqrt{\sigma_\beta}n)}
   {\beta\sqrt{\sigma_\beta}n}.
\label{eq:(A6)}
\end{equation}
Similarly, the integration in~Eq.~\eqref{eq:(A5)} yields
\begin{equation}
   \frac{1}{\lambda_0^{\text{DR}}}
   =\frac{1}{2\pi}
   \left[\frac{1}{\epsilon}
   -\frac{1}{2}\ln\left(\frac{\mathrm{e}^\gamma\sigma_\beta}{4\pi}\right)
   \right]
   +\frac{1}{\pi}\sum_{n=1}^\infty K_0(\beta\sqrt{\sigma_\beta}n).
\label{eq:(A7)}
\end{equation}
Plugging this into~Eq.~\eqref{eq:(A6)}, by noting the identity
$K_0(z)-K_2(z)=-(2/z)K_1(z)$, we have
\begin{equation}
   f(\beta)=\beta\left[
   \frac{N}{8\pi}\left(\sigma_\beta-\sigma\right)
   -\frac{N}{2\pi}\sigma_\beta\sum_{n=1}^\infty K_2(\beta\sqrt{\sigma_\beta}n)
   \right],
\label{eq:(A8)}
\end{equation}
where we have shifted the origin of the free energy density
by~$-\beta(N/8\pi)\sigma$, so that it vanishes at zero temperature
as~$\lim_{\beta\to\infty}f(\beta)/\beta=0$; note that
$\lim_{\beta\to\infty}\sigma_\beta=\sigma$
and~$\lim_{\beta\to\infty}\sum_{n=1}^\infty K_2(\beta\sqrt{\sigma_\beta}n)=0$. Since
the pressure~$P$ is related to the free energy density as~$P=-f(\beta)/\beta$
in the thermodynamic limit, we have
\begin{equation}
   P=-\frac{N}{8\pi}\left(\sigma_\beta-\sigma\right)
   +\frac{N}{2\pi}\sigma_\beta\sum_{n=1}^\infty K_2(\beta\sqrt{\sigma_\beta}n).
\label{eq:(A9)}
\end{equation}
On the other hand, the energy density is given from the free energy density
by~$\varepsilon=\partial f(\beta)/\partial\beta$. The derivative
of~Eq.~\eqref{eq:(A8)} with respect to~$\beta$ contains
$\partial\sigma_\beta/\partial\beta$, which can be deduced from the $\beta$
derivative of~Eq.~\eqref{eq:(A7)} as
\begin{equation}
   \beta\frac{\partial\sigma_\beta}{\partial\beta}
   =-\sigma_\beta\frac{4\sum_{n=1}^\infty
   \beta\sqrt{\sigma_\beta}nK_1(\beta\sqrt{\sigma_\beta}n)}
   {1+2\sum_{n=1}^\infty
   \beta\sqrt{\sigma_\beta}nK_1(\beta\sqrt{\sigma_\beta}n)},
\label{eq:(A10)}
\end{equation}
where we have used the relation~$K_0'(z)=-K_1(z)$. Using this expression and
noting the identity~$zK_2'(z)+2K_2(z)=-zK_1(z)$, we finally obtain
\begin{equation}
   \varepsilon=
   \frac{N}{8\pi}\left(\sigma_\beta-\sigma\right)
   +\frac{N}{2\pi}\sigma_\beta\sum_{n=1}^\infty K_2(\beta\sqrt{\sigma_\beta}n).
\label{eq:(A11)}
\end{equation}
Comparing Eqs.~\eqref{eq:(A9)} and~\eqref{eq:(A11)} with~Eq.~\eqref{eq:(5.24)}
given by~Eqs.~\eqref{eq:(5.21)} and~\eqref{eq:(5.22)}, we find that our lattice
energy--momentum tensor in the continuum limit correctly reproduces those
thermodynamic quantities.

\section{Naive lattice energy--momentum tensor}
\label{sec:B}
It is interesting to see how the following ``naive'' energy--momentum
tensor,\footnote{If one also applies the Noether method to the ``measure term''
(Eq.~(2.16) of~Ref.~\cite{Makino:2014sta}), the energy--momentum tensor would
have an additional term,
$-(1/2)\delta^2(0)\delta_{\mu\nu}\ln[1-\sum_{i=1}^{N-1}n^i(x)n^i(x)/N]$.
This term, however, gives rise to only sub-leading contributions in the
large-$N$ limit and does not affect the following result.}
\begin{equation}
   T_{\mu\nu}^{\text{naive}}(x)
   =\frac{1}{\lambda_0}
   \left[\partial_\mu n^i(x)\partial_\nu n^i(x)
   -\frac{1}{2}\delta_{\mu\nu}\partial_\rho n^i(x)\partial_\rho n^i(x)\right],
\label{eq:(B1)}
\end{equation}
when used in conjunction with lattice regularization, fails to reproduce the
correct answer.

Using the propagator~\eqref{eq:(3.10)}, the thermal expectation value
of~Eq.~\eqref{eq:(B1)} is given by
\begin{equation}
   \left\langle T_{00}^{\text{naive}}(x)\right\rangle_\beta
   =-\left\langle T_{11}^{\text{naive}}(x)\right\rangle_\beta
   =\frac{N}{2}\frac{1}{\beta}\sum_{-\pi/a<\omega_n<\pi/a}
   \int_{-\pi/a}^{\pi/a}\frac{\mathrm{d}p_1}{2\pi}\,
   \frac{\Hat{\omega_n}^2-\Hat{p}_1^2}
   {\Hat{\omega_n}^2+\Hat{p_1}^2+\sigma_\beta}.
\label{eq:(B2)}
\end{equation}
In this expression, we use the identity
\begin{equation}
   \frac{1}{\beta}\sum_{-\pi/a<\omega_n<\pi/a} F(\omega_n)
   =\sum_{n=-\infty}^\infty\int_{-\pi/a}^{\pi/a}\frac{\mathrm{d}p_0}{2\pi}\,
   \mathrm{e}^{ip_0\beta n}
   F(p_0)
\label{eq:(B3)}
\end{equation}
to transform the sum over~$\omega_n$ into the integral over~$p_0$. Then only
the $n=0$~term,
\begin{equation}
   \frac{N}{2}\int_p\,
   \frac{\Hat{p}_0^2-\Hat{p}_1^2}{\Hat{p}^2+\sigma_\beta},
\label{eq:(B4)}
\end{equation}
may potentially be UV divergent for~$a\to0$, but actually this term vanishes
because of the hypercubic symmetry. Other $n\neq0$ terms are UV convergent and
we may remove the lattice regulator. In this way, we have
\begin{equation}
   \left\langle T_{00}^{\text{naive}}(x)\right\rangle_\beta
   =-\left\langle T_{11}^{\text{naive}}(x)\right\rangle_\beta
   =N\sum_{n=1}^\infty
   \int\frac{\mathrm{d}^2p}{(2\pi)^2}\,\mathrm{e}^{ip_0\beta n}\,
   \frac{p_0^2-p_1^2}{p^2+\sigma_\beta}
   =-\frac{N}{2\pi}\sigma_\beta\sum_{n=1}^\infty K_2(\beta\sqrt{\sigma_\beta}n).
\label{eq:(B5)}
\end{equation}
This reproduces the expectation value of the traceless part
$\langle\{T_{00}\}_R(x)-\{T_{11}\}_R(x)\rangle_\beta$ correctly, but it misses
the trace part~$\langle\{T_{00}\}_R(x)+\{T_{11}\}_R(x)\rangle_\beta=%
-N/(4\pi)(\sigma_\beta-\sigma)$. This failure for the ``trace anomaly'' is
expected, because the naive expression~\eqref{eq:(B1)} is traceless for~$D=2$
and it cannot reproduce the trace anomaly when lattice regularization in~$D=2$
is used. Our universal formula~\eqref{eq:(5.2)} can, on the other hand,
incorporate the effect of the trace anomaly correctly, even with lattice
regularization.

\end{document}